\documentclass[12pt]{article}

\usepackage{amsmath,amsfonts}
\usepackage{amssymb}
\usepackage{amsthm}

\usepackage{times,txfonts}

\newtheorem{thm}{Theorem}[section]
\newtheorem{prop}[thm]{Proposition}
\newtheorem{cor}[thm]{Corollary}
\newtheorem{lemma}[thm]{Lemma}
\newtheorem{dfn}[thm]{Definition}

\newtheorem{remark}[thm]{\it Remark}
\newtheorem{example}[thm]{\it Example}

\numberwithin{equation}{section}

\def\pf{\noindent{\it Proof.} \ }

\def\qed{\hfill $\square$}

\def\id{{\rm id}}

\title{Universal character and $q$-difference   \\
Painlev\'e equations with affine Weyl groups}

\author{Teruhisa TSUDA}
\date{}

\begin{document}
\maketitle

\begin{abstract}
\noindent
The universal character is a polynomial 
attached to a pair of partitions and is 
a generalization of the Schur polynomial. 
In this paper, 
we introduce an integrable system of
$q$-difference lattice equations satisfied by the universal character,
and call it the {\it lattice $q$-UC hierarchy}.
We regard it as generalizing 
both $q$-KP and $q$-UC hierarchies.
Suitable similarity and periodic reductions of the hierarchy 
yield the $q$-difference Painlev\'e equations of types
$A_{2g+1}^{(1)}$ $(g \geq 1)$, $D_5^{(1)}$, and $E_6^{(1)}$.
As its consequence,
a class of algebraic solutions of the $q$-Painlev\'e equations
is rapidly obtained 
by means of 
the universal character.
In particular, 
we 
demonstrate
explicitly  the reduction procedure 
for the case of type $E_6^{(1)}$, 
via the framework of $\tau$-functions
based on the geometry of  certain rational surfaces.
\end{abstract}

\renewcommand{\thefootnote}{\fnsymbol{footnote}}

\newpage

\section{Introduction}

The present article is 
aimed to clarify the underlying relationship between 
the  universal character
and
the $q$-difference
Painlev\'e equations 
from
the viewpoint of 
infinite  integrable systems.

The universal character
$S_{[\lambda,\mu]}$, 
defined by K.~Koike \cite{k},
is a polynomial 
attached to a pair of partitions $[\lambda,\mu]$
and is a generalization of the Schur polynomial
$S_{\lambda}$.
The  universal character
describes
the character of an irreducible rational representation
of
the general linear group,
while the Schur polynomial,
as is well-known,
does 
that of an irreducible polynomial representation;
see \cite{k}, for details.

The algebraic theory of the KP hierarchy
of nonlinear partial-differential equations
is probably the most beautiful one 
in the field of classical  integrable systems.
It was discovered by M. Sato that
the totality of solutions of the KP hierarchy forms 
an infinite-dimensional Grassmann manifold;
in particular, 
the set of homogeneous polynomial solutions 
coincides with the whole set of Schur polynomials;
see  \cite{mjd,sat}.
We say that the KP hierarchy is an infinite integrable system 
which characterizes
the Schur polynomials.
On the other hand, 
an extension of the KP hierarchy
called the {\it UC hierarchy}
was proposed by the author \cite{t1};
it is  an infinite  integrable  system 
characterizing
the universal characters
as its homogeneous polynomial solutions
(see the table below).

 \begin{center}
 \begin{tabular}{ccc}
 \hline
 Character polynomials & 
 versus 
 & Infinite integrable systems
 \\
 \hline
 Schur polynomial $S_{\lambda}$
 &&
 KP hierarchy 
 \\
 $\cap$ &&$\cap$
 \\
 Universal character $S_{[\lambda,\mu]}$
 &&
 UC hierarchy  \\
 \hline 
 \end{tabular}
 \end{center}

In this paper,
we first introduce
an integrable 
system of 
$q$-difference equations
defined
on two-dimensional lattice,
called the {\it lattice  $q$-UC hierarchy}
(see Definition~\ref{def:l-q-uc}).
It is considered as generalizing both 
$q$-KP
and $q$-UC hierarchies,
which 
are the $q$-analogues of the KP and UC hierarchies;
{\it cf.}  \cite{kny2}
and  \cite{t3}
(see Remark~\ref{remark:q-uc}).
Next  we show that suitable similarity and periodic reductions of the 
lattice $q$-UC hierarchy yield the $q$-Painlev\'e equations
with affine Weyl group symmetries.
Let us refer each of $q$-Painlev\'e equations
by the Dynkin diagram of associated root system;
for example,
the $q$-Painlev\'e VI equation is represented by $D_5^{(1)}$;
see \cite{js,s}.
Then  our main result is stated as follows:

\begin{thm}  \label{thm:intro}
The $q$-Painlev\'e equations of types
$A_{2g+1}^{(1)}$ $(g \geq 1)$, $D_5^{(1)}$, and $E_6^{(1)}$
can be obtained as certain similarity reductions of the lattice $q$-UC hierarchy
with the periodic conditions of order 
$(g+1,g+1)$, $(2,2)$, and $(3,3)$, respectively.
\end{thm}

We shall demonstrate the proof of the above theorem
in detail,
particularly
for the  case of type $E_6^{(1)}$;
the other cases are briefly studied 
in Appendix.

Recall 
that the (higher order) 
$q$-Painlev\'e equation of type
$A_{N-1}^{(1)}$ is a further  generalization of 
$q$-Painlev\'e IV and V equations
which correspond to the cases $N=3$ and $4$, 
respectively;
see \cite{kny1,m}.
As shown in \cite{kny2},
it can also be obtained as 
a similarity reduction of the $q$-KP hierarchy 
with $N$-periodicity.
With this fact in mind,
we summarize
in the following table
how  the $q$-Painlev\'e equations 
relate to 
the similarity reductions of $q$-KP or lattice $q$-UC hierarchies
with periodic conditions:

\begin{center}
 \begin{tabular}{ccccc}
 \hline
 $q$-Painlev\'e equation  & $A_{2g}^{(1)}$ &   $A_{2g+1}^{(1)}$ & $D_5^{(1)}$  & $E_6^{(1)}$  
 \\ 
 \hline
 $q$-KP  hierarchy  & $2g+1$  & $2g+2$ & -- & --
 \\  
 Lattice $q$-UC  hierarchy & -- & $(g+1,g+1)$ & $(2, 2)$ & $(3, 3)$ 
 \\ 
 \hline 
 \end{tabular}
 \end{center}

The universal characters are homogeneous solutions of 
 the lattice $q$-UC hierarchy
(see Proposition~\ref{prop:uc}).
Hence we have immediately from Theorem~\ref{thm:intro}
a class of
algebraic solutions of the
$q$-Painlev\'e equations in terms of the universal character.

\begin{cor}
The $q$-Painlev\'e equations of types
$A_{2g+1}^{(1)}$ $(g \geq 1)$, $D_5^{(1)}$, and $E_6^{(1)}$
admit a class of algebraic solutions expressed in terms of the universal characters
attached  to  
pairs of $(g+1)$-, 
$2$-, and $3$-core partitions,
respectively.
\end{cor}

\begin{remark}  \rm
(i)  In  K.~Kajiwara {\it et al.} \cite{kny2}, 
rational solutions of  the $q$-Painlev\'e equations of type
$A_{N-1}^{(1)}$
were constructed by means  of the Schur polynomial
attached to an $N$-core partition,
via the similarity reduction of the $q$-KP hierarchy. 
\\
(ii) We investigated
certain similarity reductions of the $q$-UC hierarchy 
and already obtained
the same class of solutions as above
for the cases $A_{2g+1}^{(1)}$ and $D_5^{(1)}$;
see \cite{t3} and \cite{tm}.
Also, 
for $A_{3}^{(1)}$
(the $q$-Painlev\'e V equation),
the rational solutions were firstly found 
by T.~Masuda \cite{m}
without concerning any relationship to 
the infinite integrable systems.
\\
(iii) It is still an interesting open problem to obtain
the $q$-Painlev\'e equations of types $E_7^{(1)}$ and $E_8^{(1)}$
as reductions of some integrable hierarchies such as 
KP, UC, or beyond.
\end{remark}

In Section~\ref{sect:uc}, 
we introduce the lattice $q$-UC hierarchy,
which is an integrable system of $q$-difference  lattice equations
satisfied by the universal characters
(Definition~\ref{def:l-q-uc} and Proposition~\ref{prop:uc}).
In Section~\ref{sect:q-P},
we present
a  birational representation of affine Weyl group of type $E_6^{(1)}$
defined over the field of {\it $\tau$-functions},
starting from a certain configuration of nine points in the complex projective plane
(Theorem~\ref{thm:act-on-tau}).
Then we define the $q$-Painlev\'e equation of  type $E_6^{(1)}$
($q$-$P(E_6)$)
by means of the translation part of the affine Weyl group
(Definition~\ref{dfn:q-P}).
Section~\ref{sect:bil} concerns 
the system of bilinear equations  satisfied by $\tau$-functions
(Proposition~\ref{prop:bil}).
In Section~\ref{sect:sim},
we show that the bilinear form of $q$-$P(E_6)$
coincides with a similarity reduction of the lattice $q$-UC hierarchy.
 Consequently,  
 in Section~\ref{sect:alg},
 we have
 a class of algebraic solutions of $q$-$P(E_6)$ in terms of the 
 universal character
 (Theorem~\ref{thm:alg}).
 Section~\ref{sect:ver} 
  is devoted to  the proof of Proposition~\ref{prop:uc}.
 We briefly sum up in Appendix
 results on the reductions 
 to the $q$-Painlev\'e equations of  types $A_{2g+1}^{(1)}$ and $D_5^{(1)}$.
 \\

\noindent
{\it Note.} \  
Throughout this paper, 
we shall use the following convention
of
{\it $q$-shifted factorials}:
\[
(a;q)_\infty=\prod_{i=0}^{\infty}(1-a q^i),
\quad
(a;p,q)_\infty=\prod_{i,j=0}^{\infty}(1-a p^iq^j),
\]
and also
$(a_1,\ldots,a_r;q)_\infty=(a_1;q)_\infty  \cdots (a_r;q)_\infty$.

\newpage

\section{Universal characters and lattice $q$-UC hierarchy}
\label{sect:uc}

\subsection{Universal characters}

For a pair of sequences of integers
$\lambda = (\lambda_1,\lambda_2, \ldots,\lambda_l)$
and
$\mu =(\mu_1,\mu_2, \ldots,\mu_{l'}) $, 
the {\it universal character} 
$S_{[\lambda,\mu ]}({\boldsymbol x},{\boldsymbol y})$
is a polynomial in 
$({\boldsymbol x},{\boldsymbol y})=(x_1,x_2,\ldots,y_1,y_2,\ldots)$
defined 
by the determinant formula of
 {\it twisted} Jacobi--Trudi type (see \cite{k}):
\begin{equation}  \label{eq:def-of-uc}
S_{[\lambda,\mu ]}({\boldsymbol x},{\boldsymbol y})
= \det 
\left(
  \begin{array}{ll}
 p_{\mu_{l'-i+1}  +i - j }({\boldsymbol y}),  &  1 \leq i \leq l'  \\
 p_{\lambda_{i-l'}-i+j}({\boldsymbol x}),     &  l'+1 \leq i \leq l+l'   \\
  \end{array}
\right)_{1 \leq i,j \leq l+l'},
\end{equation}
where   
$p_n$ 
is a polynomial defined by 
the generating function:
\begin{equation}  
\sum_{k  \in {\mathbb Z}} p_k({\boldsymbol x})z^k = 
\exp 
\left(  \sum_{n=1}^\infty x_n z^n  \right).
\end{equation}
Schur polynomial 
$S_\lambda({\boldsymbol x})$ 
(see \cite{mac}) 
is regarded as
a special case of the universal character: 
\[
S_\lambda({\boldsymbol x}) 
= \det \bigl( p_{\lambda_i-i+j}({\boldsymbol x}) \bigr)
= S_{[\lambda, \emptyset ]}({\boldsymbol x},{\boldsymbol y}).
\]
If we count the degree of variables as
$\deg x_n=n$ and
$\deg y_n=-n$,
then the universal character
$S_{[\lambda,\mu ]}({\boldsymbol x},{\boldsymbol y})$ 
is a weighted homogeneous polynomial of degree 
$|\lambda| -|\mu|$,
where $|\lambda|=\lambda_1+\cdots+\lambda_l$.
Namely, we have 
\begin{equation}  \label{eq:hom}
S_{[\lambda,\mu ]}(c x_1,c^2 x_2,\ldots,c^{-1}y_1,c^{-2}y_2,\ldots)
=
c^{ |\lambda| -|\mu|  }
S_{[\lambda,\mu ]}(x_1,x_2,\ldots,y_1,y_2,\ldots),
\end{equation}
for any nonzero constant $c$.

\begin{example}\rm
When $\lambda=(2,1)$ and $\mu=(1)$, the universal character is 
given by
\[
S_{[(2,1),(1)]}({\boldsymbol x},{\boldsymbol y})
=
\left|
\begin{array}{lll}
p_1({\boldsymbol y}) &p_0({\boldsymbol y}) &p_{-1}({\boldsymbol y}) 
\\
p_1({\boldsymbol x}) &p_2({\boldsymbol x}) &p_{3}({\boldsymbol x}) \\
p_{-1}({\boldsymbol x}) &p_0({\boldsymbol x}) &p_{1}({\boldsymbol x}) 
\end{array}
\right|
=\left(\frac{{x_1}^3}{3}-x_3\right)y_1-{x_1}^2.
\]
\end{example}

\subsection{Lattice $q$-UC hierarchy}

Let $I\subset {\mathbb Z}_{>0}$ and $J\subset {\mathbb Z}_{<0}$
be finite indexing sets and
$t_i$ $(i \in I \cup J)$ the independent variables.
Let $T_i=T_{i;q}$ be the $q$-shift operator defined by
\[
T_{i;q}(t_i) = \left\{ 
\begin{array}{ll}
q t_i & (i \in I), \\
q^{-1} t_i &(i \in J),
\end{array} 
\right.
\]
and $T_{i;q}(t_j)=t_j$ $(i \neq j)$.
We use also the notation: 
$T_{i_1}T_{i_2} \cdots T_{i_n}=T_{i_1 i_2 \ldots i_n}$,
for the sake of brevity.

\begin{dfn}  \label{def:l-q-uc}  \rm
The following system of $q$-difference equations 
for unknowns 
$\sigma_{m,n}({\boldsymbol t})$
$(m,n \in {\mathbb Z})$
is called the {\it lattice $q$-UC hierarchy}:
\begin{equation}  \label{eq:l-q-uc}
t_i
T_{ i} (\sigma_{m,n+1})
T_{ j } (\sigma_{m+1,n}) 
-t_j
T_{ j }  (\sigma_{m,n+1})
T_{ i } (\sigma_{m+1,n}) 
=(t_i-t_j)
T_{ij}(\sigma_{m,n})
\sigma_{m+1,n+1} ,
\end{equation}
where
$i,j \in I \cup J$.
\end{dfn}

Let us consider the change of variables
\begin{equation}  \label{eq:cha}
x_n = \frac{\sum_{i \in I} {t_i}^n - q^n \sum_{j \in J}{t_j}^n }{n(1-q^n)}, 
\quad
y_n = \frac{\sum_{i \in I} {t_i}^{-n} - q^{-n} \sum_{j \in J}{t_j}^{-n} }{n(1-q^{-n})},
\end{equation}
then define the symmetric function 
$s_{[\lambda,\mu]}=s_{[\lambda,\mu]}({\boldsymbol t})$ in $t_i$ $(i \in I \cup J)$ 
by
\begin{equation}
s_{[\lambda,\mu]}({\boldsymbol t})=S_{[\lambda,\mu]}({\boldsymbol x},{\boldsymbol y}).
\end{equation}
The universal characters solve the lattice $q$-UC hierarchy
in the following sense.

\begin{prop}  \label{prop:uc}
We have
\begin{align}  
&t_i
T_{ i} (s_{[\lambda,(k',\mu)]} )
T_{ j } (s_{[(k,\lambda),\mu]}) 
-t_j
T_{ j } (s_{[\lambda,(k',\mu)]} )
T_{ i } (s_{[(k,\lambda),\mu]} )
\nonumber
\\
&=(t_i-t_j)
T_{ij}(s_{[\lambda,\mu]})
s_{[(k,\lambda),(k',\mu)]},
\label{eq:bil-of-uc}
\end{align}
for any integers $k,k'$
and sequences of integers $\lambda=(\lambda_1,\ldots,\lambda_l),\mu=(\mu_1,\ldots,\mu_{l'})$.
\end{prop}
The proof of the proposition above
will be given in Section~\ref{sect:ver}.

\begin{remark} \rm
Define the functions 
$h_n=h_n({\boldsymbol t})$ and 
$H_n=H_n({\boldsymbol t})$
by
\[
h_n({\boldsymbol t})=p_n({\boldsymbol x}),
\quad
H_n({\boldsymbol t})=p_n({\boldsymbol y}),
\]
under (\ref{eq:cha}).
We note also
the following expression 
by the generating functions:
\begin{equation}  
\sum_{k=0}^\infty h_k({\boldsymbol t}) z^k
=
 \prod_{i \in I, j \in J}
\frac{(qt_j z;q)_\infty}{(t_i z;q)_\infty}, \quad
\sum_{k=0}^\infty H_k({\boldsymbol t}) z^k
=
 \prod_{i \in I, j \in J}
\frac{(q^{-1}{t_j}^{-1} z;q^{-1})_\infty}{({t_i}^{-1} z;q^{-1})_\infty}.
\end{equation}
Hence, 
function $s_{[\lambda,\mu]}({\boldsymbol t})$
can be expressed as 
\begin{equation}  \label{eq:def-of-uc-2}
s_{[\lambda,\mu]}({\boldsymbol t})= 
\det 
\left(
  \begin{array}{ll}
 H_{\mu_{l'-i+1}  +i - j }({\boldsymbol t}),  &  1 \leq i \leq l'  \\
 h_{\lambda_{i-l'}-i+j}({\boldsymbol t}),     &  l'+1 \leq i \leq l+l'   \\
  \end{array}
\right)_{1 \leq i,j \leq l+l'}.
\end{equation}
\end{remark}

\begin{remark}  \label{remark:q-uc}
\rm  
(i) One can easily deduce from (\ref{eq:l-q-uc})
the following equation:
\begin{align}  
&(t_i-t_j)T_{ij}(\sigma_{m,n})T_k(\sigma_{m+1,n})
+(t_j-t_k)T_{jk}(\sigma_{m,n})T_i(\sigma_{m+1,n})
\nonumber
\\
&+(t_k-t_i)T_{ik}(\sigma_{m,n})
T_j(\sigma_{m+1,n})
=0,  
\label{eq:q-uc}
\end{align}
where $i,j,k \in I \cup J$,
which is exactly the bilinear equation of the $q$-UC hierarchy;
see  \cite{t3}.
\\
(ii) If $\sigma_{m,n}({\boldsymbol t})$ does not depend on $n$, that is, 
$\sigma_{m,n}= \sigma_{m,n+1}$ for all $m$ and $n$,
then (\ref{eq:l-q-uc}) is reduced to the $q$-KP hierarchy
(see \cite{kny2}):
\begin{equation}  
t_i
T_{ i} (\rho_{m})
T_{ j } (\rho_{m+1}) 
-t_j
T_{ j }  (\rho_{m})
T_{ i } (\rho_{m+1}) 
=(t_i-t_j)
T_{ij}(\rho_{m})
\rho_{m+1} ,
\end{equation}
where $\rho_m:=\sigma_{m,n}$.
\end{remark}

\section{$\tau$-functions of $q$-Painlev\'e equation}
\label{sect:q-P}

In this section we present a geometric formulation of 
the $q$-Painlev\'e equation of type $E_6^{(1)}$
by means of $\tau$-functions;
{\it cf.} \cite{s}.
Consider the configuration of nine points 
in the complex projective plane ${\mathbb P}^2$,
which are divided into three triples of collinear points.
Let $[x:y:z]$
be the homogeneous coordinate of  ${\mathbb P}^2$.
We can normalize,
without loss of generality,
the nine points 
$p_i$ $(1 \leq i \leq 9)$
under consideration as follows:
\begin{equation}
\begin{array}{l}
p_1=[0:-1:{a_3}], \quad p_2=[0:-1:{a_3}{a_6}^3], \quad 
p_3= [0:-1:{a_3} {a_6}^3 {a_0}^3], 
\\
p_4=[{a_3}:0:-1], \quad p_5=[{a_2}^3{a_3}:0:-1], \quad 
p_6= [{a_1}^3{a_2}^3{a_3}:0:-1], 
\\
p_7=[-1:{a_3}:0], \quad p_8=[-1:{a_3}{a_4}^3:0], \quad 
p_9= [-1:{a_3}{a_4}^3{a_5}^3:0],
\end{array}
\end{equation}
where
$a_i \in {\mathbb C}^\times$ 
are parameters
such that
$a_0 a_1 {a_2}^2 {a_3}^3 {a_4}^2 a_5 {a_6}^2=q$.
Let $\psi :X=X_{\boldsymbol a} \to {\mathbb P}^2$ 
be the blowing-up  
at the nine points.  
Let $e_i=\psi^{-1}(p_i)$ be the exceptional divisor
and 
$h$ the divisor class corresponding to a hyperplane.
We thus have the Picard lattice:
\[
{\rm Pic}(X)={\mathbb Z} h 
\oplus  
{\mathbb Z} e_1 \oplus \cdots \oplus {\mathbb Z} e_9, 
\] 
of rational surface $X$,
equipped with the intersection form 
$( \ | \ )$
defined by  
$(h| h)=1$,  $(e_i | e_j)=-\delta_{i,j}$ 
and
$(h| e_j)=0$.
The anti-canonical divisor 
$-K_X$
is uniquely decomposed into prime divisors:
\[
-K_X=3 h-\sum_{1 \leq i \leq 9}e_i 
= 
 D_1+D_2+ D_3,
\]
where 
$D_1=h-e_1-e_2-e_3$, 
$D_2= h-e_4-e_5-e_6$
and
$D_3=h-e_7-e_8-e_9$.
Since the dual graph of the intersections of $D_i$'s is of type $A_2^{(1)}$,
we call $X$ the {\it $A_2^{(1)}$-surface}
following the classification of the 
{\it generalized Halphen surfaces}
due to H.~Sakai \cite{s}.
The orthogonal complement
$(-K_X)^\perp  
\stackrel{\mathrm{def}}{=}
\{ v \in {\rm Pic}(X) \, | \, (v |D_i)=0 \ {\rm for} \  i=1,2,3 \}$
is isomorphic to the root lattice of type $E_6^{(1)}$.
In fact,  
$(-K_X)^\perp$ is generated by the vectors
$\alpha_{ij}=e_i-e_j$ (where both $i$ and $j$ belong to the same indexing set $\{1,2,3\}$, $\{4,5,6\}$, or $\{7,8,9\}$) and
$\alpha_{ijk}=h-e_i-e_j-e_k$ $(i\leq 3<j \leq 6< k)$;
hence
we can choose a root basis
$B=\{ \alpha_0, \ldots, \alpha_6\}$
defined by
\[
\alpha_0=\alpha_{23},\quad \alpha_1=\alpha_{56},\quad \alpha_2=\alpha_{45}, \quad
\alpha_3=\alpha_{147}, \quad 
 \alpha_4=\alpha_{78}, \quad \alpha_5=\alpha_{89},
\quad
\alpha_6=\alpha_{12},
\]
whose
Dynkin diagram is of type $E_6^{(1)}$ and looks as follows
(see, {\it e.g.},  \cite{kac}):
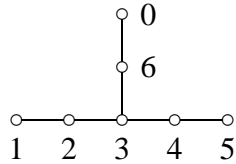
\begin{center}
\begin{picture}(200,70)

\put(60,20){\circle{4}}    \put(57,5){$1$}
\put(78,20){\line(-1,0){16}}
\put(80,20){\circle{4}}    \put(77,5){$2$}
\put(98,20){\line(-1,0){16}}
\put(100,20){\circle{4}}    \put(97,5){$3$}
\put(102,20){\line(1,0){16}}
\put(120,20){\circle{4}}    \put(117,5){$4$}
\put(122,20){\line(1,0){16}}
\put(140,20){\circle{4}}    \put(137,5){$5$}

\put(100,22){\line(0,1){16}}
\put(100,40){\circle{4}}   \put(107,36){$6$}
\put(100,42){\line(0,1){16}}
\put(100,60){\circle{4}}   \put(107,56){$0$}

\end{picture}
\end{center}
Note that the $72$ roots of $E_6$ are represented by 
 $\alpha_{ij}$ ($18$ vectors) and $\pm \alpha_{ijk}$ ($54$ vectors).
We define the action of the reflection corresponding to a root $\alpha \in (-K_X)^\perp$ by
\[
r_{\alpha}(v)=v+( v | \alpha)\alpha, \quad
v \in {\rm Pic}(X).
\]
We prepare
the notations,
$r_{ij}:=r_{\alpha_{ij}}$, $r_{ijk}:=r_{\alpha_{ijk}}$
and
$s_i:=r_{\alpha_i}$  $(i=0,\ldots,6)$,
for convenience.
Also, 
the diagram automorphism 
$\iota_i$ $(i=1,2)$ is defined by 
\[\iota_1(e_{\{1,2,3,7,8,9\}})=e_{\{7,8,9,1,2,3,\}},
\quad
\iota_2(e_{\{1,2,3,4,5,6\}})=e_{\{4,5,6,1,2,3,\}}.
\]
We thus obtain the linear action of the (extended) affine Weyl group
${\widetilde W}(E_6^{(1)})= 
\langle s_0, \ldots,s_6,\iota_1,\iota_2\rangle$
on ${\rm Pic}(X)$.
In parallel, 
we fix the action of 
$\widetilde{W}(E_6^{(1)})$
on the {\it multiplicative} 
root variables
${\boldsymbol a}=(a_0,\ldots,a_6)$
as follows:
\begin{equation} \label{eq:act-on-a}
\begin{array}{l}
s_i(a_j)=a_j a_i^{-C_{ij}}, \\
\iota_1(a_{\{0,1,2,3,4,5,6\}})={a_{\{5,1,2,3,6,0,4\}}}^{-1},
\quad
\iota_2(a_{\{0,1,2,3,4,5,6\}})={a_{\{1,0,6,3,4,5,2\}}}^{-1},
\end{array}
\end{equation}
where $(C_{ij})$ being the Cartan matrix of type $E_6^{(1)}$.

Next
we shall extend 
the linear action above  
to birational transformations.
To this end, 
we introduce the notion of $\tau$-functions;
{\it cf}. \cite{kmnoy}.
Consider the field
${\cal L}=K(\tau_1,\ldots,\tau_9)$
of rational functions in 
indeterminates $\tau_i$ $(1 \leq i \leq 9)$ 
with the coefficient field
$K={\mathbb C}({\boldsymbol a}^{1/3})={\mathbb C}({a_0}^{{1}/{3}},\ldots,{a_6}^{{1}/{3}})$.
Take a sub-lattice    
$M=\bigcup_{i=1,2,3} M_i$ 
of ${\rm Pic}(X)$,
where 
\[
M_i=\left\{v \in {\rm Pic}(X) \,  \big|  \,  (v | v)= -(v | D_i)=-1, \  (v | D_j)=0 \  (j\neq i)\right\}.
\]

\begin{dfn}   \rm
\label{dfn:tau}
A function $\tau: M  \to  {\cal L}$
is said to be a {\it $\tau$-function} iff it satisfies the conditions{\rm:} \\
{\rm (i)} $\tau(w.v)=w. \tau(v)$ for any  
$v  \in  M$
and
$w \in \widetilde{W}(E_6^{(1)})${\rm;} 
{\rm (ii)} $\tau(e_i)=\tau_i$ $(1 \leq i \leq 9)$.
\end{dfn}

Such functions
and  
the action of  
$ \widetilde{W}(E_6^{(1)})$ 
on them are explicitly determined in the following way.
Any divisor 
$\Lambda=n h -e_{i_1}-\cdots -e_{i_{n+1}} \in M$
corresponds to a curve of degree $n$ on ${\mathbb P}^2$
passing through $n+1$ points 
$p_{i_1}, \ldots,p_{i_{n+1}}$
(with counting the multiplicity). 
We can choose uniquely the {\it normalized} defining 
polynomial $F_{\Lambda}(x,y,z)=\sum_{i+j+k=n} A_{ijk} x^iy^jz^k
\in {\mathbb Q}({\boldsymbol a})[x,y,z]$
of the curve,
such that 
$\prod A_{ijk}=1$.
For example, 
we have
\begin{align*}
F_{h-e_1-e_4}&= {a_3}^{-1}  x+ {a_3}  y+ z,
\\
F_{h-e_4-e_7}&= x+ {a_3}^{-1}  y+ {a_3}   z,
\\
F_{h-e_1-e_7}&= {a_3}  x+y+ {a_3}^{-1}   z.
\end{align*}
Let
\begin{equation}   \label{eq:xyz}
\left(\frac{x}{c_{x}}, \frac{y}{c_{y}},\frac{z}{c_{z}}  \right)
=(\tau_1\tau_2\tau_3, \tau_4\tau_5\tau_6, \tau_7\tau_8\tau_9 ),
\end{equation}
where
\[
c_{x}={a_1}^{\frac{1}{3}}{a_2}^{\frac{2}{3}}{a_4}^{-\frac{2}{3}}{a_5}^{-\frac{1}{3}},
\quad
c_{y}={a_5}^{\frac{1}{3}}{a_4}^{\frac{2}{3}}{a_6}^{-\frac{2}{3}}{a_0}^{-\frac{1}{3}},
\quad
c_{z}={a_0}^{\frac{1}{3}}{a_6}^{\frac{2}{3}}{a_2}^{-\frac{2}{3}}{a_1}^{-\frac{1}{3}}.
\]
Suppose that
\begin{equation}  \label{eq:Fijk}
F_{\Lambda}(x,y,z)=\tau(n h-e_{i_1}- \cdots-e_{i_{n+1}})
\tau(e_{i_1}) \cdots \tau(e_{i_{n+1}}).
\end{equation}
Therefore
we see that
the linear action of 
$\widetilde{W}(E_6^{(1)})$
on $M$ yields
the action on $\tau$-functions immediately.
For instance,
by using $r_{ijk}(e_k) = h-e_i-e_j$, 
we can compute
the action of $r_{ijk}$:
\[
r_{ijk}(\tau(e_k))= \tau(h-e_i-e_j)=\frac{F_{h-e_i-e_j}(x,y,z)}{\tau(e_i)\tau(e_j)}.
\]
Each action of $r_{ij}$ and diagram automorphism $\iota_i$ is realized as just a permutation of
$\tau_i$'s.
Summarizing above, we now arrive at the following theorem.

\begin{thm}   \label{thm:act-on-tau}
Define the birational transformations 
$s_i$ $(0 \leq i\leq 6)$ and $\iota_j$ $(j=1,2)$ on
${\cal L}={\mathbb C}({\boldsymbol a}^{1/3})(\tau_1,\ldots,\tau_9)$
by
\begin{equation}  \label{eq:act-on-tau}
\begin{array}{l}
s_1(\tau_{\{5,6\}})=\tau_{\{6,5\}}, 
\quad 
s_2(\tau_{\{4,5\}})=\tau_{\{5,4\}}, 
\quad
s_4(\tau_{\{7,8\}})=\tau_{\{8,7\}}, 
\quad
s_5(\tau_{\{8,9\}})=\tau_{\{9,8\}}, 
\\ 
s_6(\tau_{\{1,2\}})=\tau_{\{2,1\}}, 
\quad 
s_0(\tau_{\{2,3\}})=\tau_{\{3,2\}}, 
\quad
\iota_1(\tau_{\{1,2,3\}})=\tau_{\{7,8,9\}}, 
\quad
\iota_2(\tau_{\{1,2,3\}})=\tau_{\{4,5,6\}}, 
\\
s_3(\tau_1)=\left(  
c_{x}\tau_1\tau_2\tau_3
+   {a_3}^{-1}c_{y}\tau_4\tau_5\tau_6
+ {a_3}c_{z}\tau_7\tau_8\tau_9
\right)/({\tau_4}{\tau_7}), 
\\
s_3(\tau_4)=\left(  
{a_3} c_{x}\tau_1\tau_2\tau_3
+   c_{y}\tau_4\tau_5\tau_6
+ {a_3}^{-1}c_{z}\tau_7\tau_8\tau_9
\right)/({\tau_1}{\tau_7}), 
\\
s_3(\tau_7)=\left(  
{a_3}^{-1} c_{x}\tau_1\tau_2\tau_3
+  {a_3} c_{y}\tau_4\tau_5\tau_6
+ c_{z}\tau_7\tau_8\tau_9
\right)/({\tau_1}{\tau_4}).
\end{array}
\end{equation}
Then {\rm(\ref{eq:act-on-tau})} with {\rm(\ref{eq:act-on-a})}
provide a realization of 
$\widetilde{W}(E_6^{(1)})=\langle s_0,\ldots,s_6,\iota_1,\iota_2 \rangle$.
\end{thm}

Let 
\begin{equation}  \label{eq:fg}
[f:g:1]
=\left[\frac{x}{c_{x}}: \frac{y}{c_{y}}:\frac{z}{c_{z}}  \right]
=\left[\tau_1\tau_2\tau_3: \tau_4\tau_5\tau_6: \tau_7\tau_8\tau_9 \right].
\end{equation}
By virtue of Theorem~\ref{thm:act-on-tau},
we obtain the following birational transformations 
on the inhomogeneous coordinate $(f,g)$:
\begin{equation}\label{eq:act-on-(f,g)}
\begin{array}{l}
\displaystyle
s_3(f)=f \frac{ c_{x}f+{a_3}^{-1}  c_{y} g+ {a_3}  c_{z}  }{
{a_3}^{-1}  c_{x} f+{a_3}   c_{y} g+c_{z}
},
\\
\displaystyle
s_3(g)=g \frac{ {a_3} c_{x} f+ c_{y} g+ {a_3}^{-1}  c_{z}  }{
{a_3}^{-1}  c_{x} f+{a_3}   c_{y} g+c_{z} },
\\ 
\displaystyle
\iota_1(f)=\frac{1}{f}, \quad \iota_1(g)=\frac{g}{f}, 
\quad
\iota_2(f)=g,  \quad \iota_2(g)=f.
\end{array}
\end{equation}
The birational action
arising from 
the translation part of 
affine Weyl group can be regarded
as a discrete dynamical system 
and is called a  discrete Painlev\'e equation;
{\it cf}. \cite{ny}.
Consider an element 
\begin{equation}
\ell =  r_{258} r_{369}  r_{258}   r_{147} 
=   
( 
s_2   s_4  s_6  s_0  s_1  s_5  s_3  s_2  s_4  s_6  s_3
)^2
 \in
W(E_6^{(1)}),
\end{equation}
acting on the parameters 
${\boldsymbol a}=(a_0,\ldots,a_6)$
as their $q$-shifts:
\begin{equation}  
\ell({\boldsymbol a})
=\overline{{\boldsymbol a}}
=(a_0,a_1, q^{-1} a_2, q^{2}a_3, q^{-1} a_4,a_5,q^{-1} a_6).
\end{equation}
We define rational functions 
$F({\boldsymbol a};f,g), \  G({\boldsymbol a};f,g) \in {\mathbb C}({\boldsymbol a}^{1/3};f,g)$
by
\begin{equation}
\ell(f)=F({\boldsymbol a};f,g), 
\quad
\ell(g)=G({\boldsymbol a};f,g).
\end{equation}
\begin{dfn}  \label{dfn:q-P}
\rm
The system of functional equations
\begin{equation}\label{eq:q-P}
f(\overline{{\boldsymbol a}})=F({\boldsymbol a};f({\boldsymbol a}),g({\boldsymbol a})),
\quad
g(\overline{{\boldsymbol a}})=G({\boldsymbol a};f({\boldsymbol a}), g({\boldsymbol a})),
\end{equation}
for unknowns 
$f=f({\boldsymbol a})$ and $g=g({\boldsymbol a})$
is called the {\it $q$-Painlev\'e equation of type $E_6^{(1)}$}.
\end{dfn}

We shall often denote  
(\ref{eq:q-P}) 
shortly by $q$-$P(E_6)$.

\begin{remark}\rm
\label{remark:weyl}
We have the following inclusion relation of affine Weyl groups:
$W(E_6^{(1)})\supset
W(A_5^{(1)})  \oplus W(A_1^{(1)})$.
For instance, 
the sets of vectors
$B'=\{  \alpha_{158}, \alpha_{367}, \alpha_{248}, \alpha_{169}, \alpha_{257}, \alpha_{349} 
\}$   and    
$B''=\{
\alpha_{147}, \alpha_{258}+\alpha_{369}\}$
realize the root bases of types 
$A_5^{(1)}$ and $A_1^{(1)}$,
respectively.
Moreover,
they
are mutually orthogonal.
The transformation $\ell$, 
used to define the $q$-Painlev\'e equation (\ref{eq:q-P}),
is exactly the translation in $W(A_1^{(1)})$;
that is,
$r_{\alpha_{258}+\alpha_{369}}  
r_{\alpha_{147}} 
= (r_{258} r_{369} r_{258} ) r_{147} =\ell$.
\end{remark}

\section{Bilinear equations among $\tau$-functions}
\label{sect:bil}

Let us 
introduce the transformations
$\ell_2 = r_{369}  r_{147}  r_{369}  r_{258}$ and
$\ell_3 = r_{147}  r_{258}  r_{147}  r_{369}$,
in parallel with 
$\ell_1=\ell=
r_{258} r_{369} r_{258}  r_{147} $.
These act on the root variables as their $q$-shifts:
\begin{equation}
\begin{array}{l}
\ell_1({\boldsymbol a})
=(a_0,a_1, q^{-1} a_2, q^{2}a_3, q^{-1} a_4,a_5,q^{-1} a_6),
\\
\ell_2({\boldsymbol a})
=(q^{-1} a_0, q^{-1} a_1, q  a_2, q^{-1} a_3,  q a_4,q^{-1} a_5,q a_6),
\\
\ell_3({\boldsymbol a})
=(q a_0,q a_1, a_2, q^{-1} a_3,  a_4,q a_5, a_6).
\end{array}
\end{equation}
Note that $\ell_i$'s are mutually commutable and $\ell_1 \ell_2  \ell_3=\id$.
The action of $\ell_i$ on the auxiliary variables
\begin{equation}
a=(a_0a_1a_5)^{1/3}, \quad 
b=(a_2a_4a_6 q)^{1/3},
\end{equation}
is described as follows:
\begin{equation}  \label{eq:ell-ab}
\ell_1(a,b)=(a,q^{-1}b),  \quad
\ell_2(a,b)=(q^{-1}a,q b),  \quad
\ell_3(a,b)=(qa,b).
\end{equation}

\begin{lemma}  \label{lemma:bil-lemma}
It holds that
\begin{equation}  \label{eq:bil-lemma}
\tau_3 \ell_3(\tau_6) 
-a^2b \ell_3(\tau_3)\tau_6
=
\left( 
\frac{{a_1}^2a_2}{{a_0}^2a_6}
\right)^{1/3}
\frac{1-a^6b^3}{a^2b}
\tau_7\tau_8.
\end{equation}
\end{lemma}

\pf
We have
(see Section~\ref{sect:q-P})
\begin{align*}
F_{h-e_3-e_9}(x,y,z)
&=
{a_0}a_3{a_4}^2{a_5}^2{a_6}x
+\frac{a_0a_6 y}{a_4a_5}
+\frac{z}{{a_0}^{2}a_3a_4a_5{a_6}^{2}}, \\
F_{h-e_6-e_9}(x,y,z)
&=
\frac{a_4a_5x}{a_1a_2}
+\frac{y}{a_1a_2a_3{a_4}^2{a_5}^2}
+{a_1}^{2}{a_2}^{2}{a_3}{a_4}{a_5}z.
\end{align*}
Eliminating $x$ and $y$, we get
\begin{equation}
\label{eq:f39-f69}
F_{h-e_3-e_9}-a_0a_1a_2a_3a_4a_5a_6 F_{h-e_6-e_9}=
\frac{1-(a_0a_1a_2a_3a_4a_5a_6)^3}{{a_0}^2a_3a_4a_5{a_6}^2} z.
\end{equation}
Recall that $z=c_{z}\tau_7\tau_8\tau_9$
and $F_{h-e_i-e_j}=\tau_i \tau_j\tau(h-e_i-e_j)$.
By virtue of 
${\ell_3}(e_6)=h-e_3-e_9$
and
${\ell_3}(e_3)=h-e_6-e_9$,
we thus obtain (\ref{eq:bil-lemma}) from (\ref{eq:f39-f69}).
\qed
\\

We shall rename the $\tau$-functions
as follows:
\begin{equation}  \label{eq:def-UVW}
U_{\{1,2,3\}} 
=\frac{ \tau_{\{1,4,7\}} }{N(a,b)},
\quad
V_{\{1,2,3\}}
=\frac{ \tau_{\{2,5,8\}} }{N(q^{1/3}a,q^{-2/3}b)},
\quad
W_{\{1,2,3\}}
=\frac{ \tau_{\{3,6,9\}} }{N(q^{-1/3}a,q^{-1/3}b)},
\end{equation}
where the normalization factor $N(a,b)$ is defined by 
\begin{equation}
N(a,b)
=
\frac{   \displaystyle
\left(
-\frac{aq}{b},-ab^2q,-\frac{q}{a^2b};
q,q
\right)_\infty
\left(
\frac{b^3q^3}{a^3},\frac{q^3}{a^3b^6},a^6b^3q^3;
q^3,q^3
\right)_\infty
}{   \displaystyle
\left(
\frac{b^2q^2}{a^2},\frac{q^2}{a^2b^4},a^4b^2q^2;
q^2,q^2
\right)_\infty
}.
\end{equation}
Equation (\ref{eq:bil-lemma}) in Lemma~\ref{lemma:bil-lemma}
is then rewritten into 
\begin{equation}    \label{eq:pre-bil}
\frac{1}{a}
W_1 \ell_3(W_2) 
-a b \ell_3(W_1) W_2
=
\left( 
\frac{{a_1}^2a_2}{{a_0}^2a_6}
\right)^{1/3}
\left(
   \frac{1}{a} -ab
 \right)
U_3 V_3,
\end{equation}
by straightforward computation.
As seen below,
all the other bilinear equations for $U_i$, $V_i$ and $W_i$
can also be derived from (\ref{eq:pre-bil})
by suitable symmetries of ${\widetilde W}(E_6^{(1)})$.
Applying $r_{13}r_{46}r_{79}$ to  (\ref{eq:pre-bil})
and viewing that 
$\ell_1=r_{13}r_{46}r_{79}\ell_3r_{13}r_{46}r_{79}$, 
we thus obtain
\begin{equation}  \label{eq:pre-bil-2}
ab U_1 \ell_1(U_2) 
- \frac{q}{b}
 \ell_1(U_1) U_2
=
\left( 
\frac{a_0{a_6}^2}{a_1{a_2}^2}
\right)^{1/3}
\left(ab -\frac{q}{b}
 \right)
V_3 W_3.
\end{equation}
Moreover, 
we consider an element 
$\pi=
s_0s_1s_5 
\iota_1\iota_2
\in {\widetilde W}(E_6^{(1)})$ 
of order six
whose
action  is given as follows:
\[
\pi:(a_0,a_1,a_2,a_3,a_4,a_5,a_6;\tau_{ \{1,2,3,4,5,6,7,8,9 \} })
\mapsto
\left(
\frac{1}{a_5},
\frac{1}{a_0},
a_0a_6,
a_3,
a_1a_2, 
\frac{1}{a_1},
a_4a_5;
\tau_{ \{7,9,8,1,3,2,4,6,5 \} }
\right).
\]
Hence we see that
\begin{equation}
\pi:
(a,b;U_i,V_i,W_i)
\mapsto
\left(\frac{1}{a},ab;
U_{i-1},W_{i-1}, V_{i-1}
\right),
\end{equation}
and also that the commutation relations
$\pi \ell_1= \ell_1  \pi$,
$\pi  \ell_2 =\ell_3  \pi$
and
$\pi  \ell_3 =\ell_2  \pi$
hold.
Note that $\pi$ realizes the rotational diagram automorphism of 
$A_5^{(1)}$, 
considered in Remark~\ref{remark:weyl}.
Applying $\pi$ to (\ref{eq:pre-bil}) and (\ref{eq:pre-bil-2}),
we  get the following proposition.

\begin{prop}   \label{prop:bil}
The following bilinear equations among the 
$\tau$-functions
$U_i$, $V_i$ and $W_i$ hold{\rm:}
\begin{subequations}  
\label{subeq:bil-UVW}
\begin{align}  \label{eq:bil-UVW-1}
ab U_i \ell_1(U_{i+1}) 
- \frac{q}{b}
 \ell_1(U_i) U_{i+1}
&=
\gamma_i
\left(ab -\frac{q}{b}
 \right)
V_{i+2} W_{i+2},
\\   \label{eq:bil-UVW-2}
\frac{1}{b} V_i \ell_2(V_{i+1})
-\frac{1}{a} \ell_2(V_i) V_{i+1}
&=
\delta_i\left (\frac{1}{b}-\frac{1}{a}\right)W_{i+2} U_{i+2},
\\   \label{eq:bil-UVW-3}
\frac{1}{a}
W_i \ell_3(W_{i+1}) 
-a b \ell_3(W_i) W_{i+1}
&=
\epsilon_{i}
\left(
  \frac{1}{a} -ab
 \right)
U_{i+2} V_{i+2},
\end{align}
\end{subequations}
for $i \in{\mathbb Z}/ 3 {\mathbb Z}$.
Here $\gamma_i$, $\delta_i$ and $\epsilon_i$
are the parameters defined by
\begin{equation}
\begin{array}{lll}
\displaystyle
\gamma_1= 
\left( 
\frac{a_0{a_6}^2}{a_1{a_2}^2}
\right)^{1/3},
&\displaystyle
\gamma_2=
\left( 
\frac{a_1{a_2}^2}{{a_4}^2a_5}
\right)^{1/3},
&\displaystyle
\gamma_3=
\left( 
\frac{{a_4}^2a_5}{a_0{a_6}^2}
\right)^{1/3},
\\
\displaystyle
\delta_1= \left(\frac{a_0a_2}{ a_1 a_6}\right)^{1/3} ,
& \displaystyle
\delta_2=
\left(\frac{a_1a_4}{ a_2 a_5}\right)^{1/3} ,
& \displaystyle
\delta_3=
\left(\frac{a_5a_6}{ a_0 a_4}\right)^{1/3} ,
\\
\displaystyle
\epsilon_1=
\left( 
\frac{{a_1}^2a_2}{{a_0}^2a_6}
\right)^{1/3},
&\displaystyle
\epsilon_2=
\left( 
\frac{a_4{a_5}^2}{{a_1}^2a_2}
\right)^{1/3},
&\displaystyle
\epsilon_3=
\left( 
\frac{{a_0}^2 a_6}{a_4{a_5}^2}
\right)^{1/3}.
\end{array}
\end{equation}
\end{prop}

We call system (\ref{subeq:bil-UVW}) the  {\it
bilinear form of the 
$q$-Painlev\'e equation of type $E_6^{(1)}$}.
Conversely, 
we can verify that,
for  any functions
$U_i,V_i,W_i$ $(i \in{\mathbb Z}/ 3 {\mathbb Z})$
satisfying (\ref{subeq:bil-UVW}), 
the pair 
$(f,g)$ defined by
\[
f=\frac{U_1V_1W_1}{U_3V_3W_3},
\quad
g=\frac{U_2V_2W_2}{U_3V_3W_3},
\]
certainly solves the 
$q$-Painlev\'e equation (\ref{eq:q-P});
here we
recall (\ref{eq:fg}) and (\ref{eq:def-UVW}).

\section{Similarity reduction of lattice $q$-UC hierarchy to $q$-$P(E_6)$}
\label{sect:sim}

We shall explain 
how the bilinear form of $q$-$P(E_6)$,
(\ref{subeq:bil-UVW}), 
arises naturally from the lattice $q$-UC hierarchy,
through  
certain periodic and similarity reductions.  
Let $I=\{1,2,3\}$ and $J=\emptyset$
and
consider the lattice $q$-UC hierarchy:
\begin{equation}    \label{eq:l-q-uc-2}
t_i
T_{ i} (\sigma_{m,n+1})
T_{ j } (\sigma_{m+1,n}) 
-t_j
T_{ j }  (\sigma_{m,n+1})
T_{ i } (\sigma_{m+1,n}) 
=(t_i-t_j)
T_{ij}(\sigma_{m,n})
\sigma_{m+1,n+1}.
\end{equation}

We impose the $(3,3)$-periodic condition:
\begin{equation}  \label{eq:percon}
\sigma_{m,n}=\sigma_{m+3,n}=\sigma_{m,n+3},
\end{equation}
and the similarity condition:
\begin{equation}   \label{eq:simcon}
\sigma_{m,n}(c t_1,ct_2,ct_3)=c^{d_{m,n}}\sigma_{m,n}(t_1,t_2,t_3),
\end{equation}
for any $c \in {\mathbb C}^\times$.
Here $d_{m,n}$ are constant parameters such that
$d_{m,n}+d_{m+1,n+1}=d_{m+1,n}+d_{m,n+1}$.
We introduce the functions 
$\widetilde{\sigma}_{m,n}(a,b)$ 
in two variables
defined by
$\widetilde{\sigma}_{m,n}(a,b)=\sigma_{m,n}(t_1,t_2,t_3)$
under the substitution
$(t_1,t_2,t_3)=(a^{-1},b^{-1},ab)$.
We thus have the following lemma.

\begin{lemma}   \label{lemma:simred}
Let
\begin{align}  \nonumber
U_i(a,b)&=\widetilde{\sigma}_{i,-i}(a,b),
\\
V_i(a,b)&=\widetilde{\sigma}_{i+1,-i+1}(q^{1/3}a,q^{-2/3}b),
\\  
\nonumber
W_i(a,b)&=\widetilde{\sigma}_{i+2,-i+2}(q^{-1/3}a,q^{-1/3}b),
\end{align}
for $i \in {\mathbb Z}/3{\mathbb Z}$.
Then these functions satisfy the bilinear form of $q$-$P(E_6)$,
{\rm(\ref{subeq:bil-UVW})},
with the parameters{\rm:}
\begin{equation}
\gamma_i=q^{(d_{i,-i+2}-d_{i+1,-i})/3},
\quad
\delta_i=q^{(d_{i+1,-i}-d_{i+2,-i+1})/3},
\quad
\epsilon_i=q^{(d_{i+2,-i+1}-d_{i,-i+2})/3}.
\end{equation}
\end{lemma}

\pf
Being attentive to the action of $\ell_i$'s 
on variables $a$ and $b$ (see (\ref{eq:ell-ab})),
one can deduce the bilinear form of $q$-$P(E_6)$
straightforwardly from the lattice $q$-UC hierarchy (\ref{eq:l-q-uc-2})
by the similarity condition (\ref{eq:simcon})
together with the periodicity (\ref{eq:percon}).

For instance,
we shall start from (\ref{eq:l-q-uc-2}) with $(m,n)=(r+1,-r)$ and $(i,j)=(1,2)$:
\begin{align*}
&t_1
\sigma_{r+1,-r+1}(q t_1,t_2,t_3)
\sigma_{r+2,-r}(t_1,q t_2,t_3)
-t_2
\sigma_{r+1,-r+1}(t_1,q t_2,t_3)
\sigma_{r+2,-r} (q t_1,t_2,t_3)
\\
&=(t_1-t_2)
\sigma_{r+1,-r}(q t_1,q t_2,t_3)
\sigma_{r+2,-r+1}(t_1,t_2,t_3).
\end{align*}
By using the homogeneity  
(\ref{eq:simcon}),
we have
\begin{align*}
&
q^{(d_{r+1,-r+1}+d_{r+2,-r})/3}
t_1
\sigma_{r+1,-r+1}(q^{2/3} t_1, q^{-1/3}t_2,q^{-1/3}t_3)
\sigma_{r+2,-r}(q^{-1/3}t_1,q^{2/3} t_2,q^{-1/3}t_3)
\\
&-q^{(d_{r+1,-r+1}+d_{r+2,-r})/3}
t_2
\sigma_{r+1,-r+1}(q^{-1/3}t_1,q^{2/3} t_2,q^{-1/3}t_3)
\sigma_{r+2,-r} (q^{2/3} t_1, q^{-1/3}t_2,q^{-1/3}t_3)
\\
&=
q^{2 d_{r+1,-r}/3}
(t_1-t_2)
\sigma_{r+1,-r}(q^{1/3} t_1,q^{1/3} t_2,q^{-2/3}t_3)
\sigma_{r+2,-r+1}(t_1,t_2,t_3).
\end{align*}
Putting $(t_1,t_2,t_3)=(a^{-1},b^{-1},ab)$,
therefore we obtain
\begin{align*}
&
\frac{1}{a}
\widetilde{\sigma}_{r+1,-r+1}(q^{-2/3}a, q^{1/3}b)
\widetilde{\sigma}_{r+2,-r}(q^{1/3}a,q^{-2/3}b)
\\
&-
\frac{1}{b}
\widetilde{\sigma}_{r+1,-r+1}(q^{1/3}a,q^{-2/3}b)
\widetilde{\sigma}_{r+2,-r}(q^{-2/3}a, q^{1/3}b) 
\\
&=
q^{(d_{r+1,-r}-d_{r+2,-r+1})/3}
\left(\frac{1}{a}-\frac{1}{b}
\right)
\widetilde{\sigma}_{r+1,-r}(q^{-1/3}a,q^{-1/3} b)
\widetilde{\sigma}_{r+2,-r+1}(a,b),
\end{align*}
which turns out to coincide with (\ref{eq:bil-UVW-2})
in view of the action of $\ell_2$.
In the same way, we can derive also (\ref{eq:bil-UVW-1}) and (\ref{eq:bil-UVW-3}).
The proof is now complete.
\qed

\section{Algebraic solutions of $q$-Painlev\'e equation in terms of the universal character}
\label{sect:alg}

As seen in the preceding section,
the $q$-Painlev\'e equation of type $E_6^{(1)}$
is in fact
equivalent to
a similarity reduction of the (periodic)
lattice $q$-UC hierarchy.
On the other hand,
we have already known that
the lattice $q$-UC hierarchy admits the universal characters 
as its homogeneous solutions;
see Proposition~\ref{prop:uc}.
Consequently, we obtain in particular
a class of
algebraic  solutions of 
the $q$-Painlev\'e equation 
in terms of the universal character.

In order to state our result precisely, 
we first recall the notion of $N$-core partitions;
see, {\it e.g.},  \cite{n}.
A subset $M \subset {\mathbb Z}$ is said to be a 
{\it Maya diagram} 
if
$m \in M$ $(m \ll 0)$ 
and
$m \notin M$ $(m \gg 0)$.
Each Maya diagram 
$M = \{ \ldots, m_3, m_2, m_1 \}$
corresponds to
a unique partition
$\lambda=(\lambda_1,\lambda_2,\ldots)$ 
such that $m_i-m_{i+1}=\lambda_i-\lambda_{i+1}+1$.
For a sequence of integers
${\boldsymbol n}=(n_1,n_2,\ldots,n_{N})\in{\mathbb Z}^{N}$,
let us consider
the Maya diagram
\[
M({\boldsymbol n})=
(N{\mathbb Z}_{<n_1}+1)
\cup
(N{\mathbb Z}_{<n_2}+2)
\cup
\cdots
\cup
(N{\mathbb Z}_{<n_N}+N),
\]
and denote by 
$\lambda({\boldsymbol n})$
the corresponding partition.
Note that 
$\lambda({\boldsymbol n})=\lambda({\boldsymbol n}+{\boldsymbol 1})$
where ${\boldsymbol 1}=(1,1,\ldots,1)$.
We call a partition of the form $\lambda({\boldsymbol n})$
an {\it $N$-core partition}.
It is well-known that a partition $\lambda$ is 
$N$-core 
if and only if
$\lambda$ 
has no hook with length of a multiple of $N$.
We have a cyclic chain of the universal characters 
attached to $N$-core partitions;
see \cite[Lemma~2.2]{t2}.

\begin{lemma}
\label{lemma:uc-c}
It holds that
\begin{equation}
S_{\left[\bigl(k_i, 
\,
\lambda({\boldsymbol n}(i-1) ) \bigr), 
\,
\mu \right]}
=
\pm S_{\left[\lambda({\boldsymbol n}(i) ), 
\,
\mu \right]},
\end{equation}
for arbitrary 
${\boldsymbol n} =(n_1,n_2,\ldots,n_N)
\in {\mathbb Z}^N$
and partition $\mu$.
Here
${\boldsymbol n}(i) ={\boldsymbol n} +(\overbrace{1,\ldots,1}^i ,\overbrace{0,\ldots,0}^{N-i})$
and
$k_i=N n_i-|  {\boldsymbol n} |$  with $|  {\boldsymbol n} |=n_1+n_2+\cdots+n_N$.
\end{lemma}

Finally,
by virtue of Proposition~\ref{prop:uc}
and
 Lemmas~\ref{lemma:simred}
 and
 \ref{lemma:uc-c},
we are led to the following 
expression of
algebraic solutions 
by means of 
the universal character
attached to 
a pair of three-core partitions.
Define a rational function 
$R_{[\lambda,\mu]}=R_{[\lambda,\mu]}(a,b)$
by 
(recall (\ref{eq:def-of-uc}) or (\ref{eq:def-of-uc-2}))
\begin{equation}
R_{[\lambda,\mu]}(a,b)
=S_{[\lambda,\mu]}({\boldsymbol x}, {\boldsymbol y})
=s_{[\lambda,\mu]}({\boldsymbol t}),
\end{equation}
under the substitution:
\begin{equation}
x_n=\frac{a^{-n}+b^{-n}+(ab)^n}{n(1-q^n)},
\quad
y_n=\frac{a^{n}+b^{n}+(ab)^{-n}}{n(1-q^{-n})},
\end{equation}
or 
$(t_1,t_2,t_3)=(a^{-1},b^{-1}, ab)$
with $I=\{1,2,3 \}$ and $J=\emptyset$.

\begin{thm}  \label{thm:alg}
For any 
${\boldsymbol m}=(m_1,m_2,m_3), {\boldsymbol n}=(n_1,n_2,n_3) \in {\mathbb Z}^3$,
let
\begin{align}
U_i(a,b)&=
R_{\left[ \lambda({\boldsymbol m}(i)) ,  
\,
\lambda({\boldsymbol n}(-i))   \right]}
(a,b),
\nonumber \\
V_i(a,b)&=
R_{\left[ \lambda({\boldsymbol m}(i+1)) ,  
\,
\lambda({\boldsymbol n}(-i+1))   \right]}
(q^{1/3}a,q^{-2/3}b),
\\
W_i(a,b)&=
R_{\left[ \lambda({\boldsymbol m}(i+2)) ,  
\,
\lambda({\boldsymbol n}(-i+2))   \right]}
(q^{-1/3}a,q^{-1/3}b).
\nonumber
\end{align}
{\rm(i)}
These functions solve the system of bilinear equations {\rm(\ref{subeq:bil-UVW})}
with the parameters{\rm:}
\begin{equation}
\gamma_i=  q^{n_{-i}-m_{i+1}+\frac{| \boldsymbol m |-| \boldsymbol n |}{3}},
\quad
\delta_i=   q^{n_{-i+1}-m_{i+2}+\frac{| \boldsymbol m |-| \boldsymbol n |}{3}},
\quad
\epsilon_i=  q^{n_{-i+2}-m_{i}+\frac{| \boldsymbol m |-| \boldsymbol n |}{3}}.
\end{equation}
{\rm(ii)}
Consequently,
the pair of functions
\begin{equation}
f= \frac{U_1V_1W_1}{U_3V_3W_3}, 
\quad 
g= \frac{U_2V_2W_2}{U_3V_3W_3},
\end{equation}
gives an algebraic solution
of the $q$-Painlev\'e equation of type $E_6^{(1)}$,
{\rm(\ref{eq:q-P})},
when
\begin{equation}
\begin{array}{lll}
a_1=  a q^{\frac{|\boldsymbol  m|+|\boldsymbol n|}{3}-m_1-n_3},
&
a_5= a q^{\frac{|\boldsymbol  m|+|\boldsymbol n|}{3}-m_2-n_2} ,
&
a_0= a q^{\frac{|\boldsymbol  m|+|\boldsymbol n|}{3}-m_3-n_1} ,
\\
a_2=  b q^{\frac{|\boldsymbol  m|+|\boldsymbol n| -1}{3}-m_3-n_2},
&
a_4=   b q^{\frac{|\boldsymbol  m|+|\boldsymbol n| -1}{3}-m_1-n_1},
&
a_6= b q^{\frac{|\boldsymbol  m|+|\boldsymbol n| -1}{3}-m_2-n_3} .
\end{array}
\end{equation}
\end{thm}

\begin{example}  \rm
Let us consider the function
\[
P_{[\lambda,\mu]}(a,b;q)
=(a b)^{|\lambda|+|\mu|}  
q^{-| \nu |}
\prod_{(i,j) \in \lambda} \left(1-q^{h(i,j)}\right)
\prod_{(k,l) \in \mu} \left(q^{h(k,l)}-1\right)
R_{[\lambda,\mu]}(a,b),
\]
associated with the algebraic solutions
given in Theorem~\ref{thm:alg}.
Here we denote by $h(i,j)$ 
the {\it hook-length},
that is,
$h(i,j)=\lambda_i+\lambda'_j-i-j+1$
(see \cite{mac})
and  let $\nu=(\nu_1,\nu_2,\ldots)$ be a sequence of integers
defined by
$\nu_i=\max\{0,\mu'_i-\lambda_i\}$.
It is interesting that 
$P_{[\lambda,\mu]}(a,b;q)$
forms a polynomial whose coefficients are all positive integers.
A few examples of the {\it special polynomials}
are given below:
\[
\begin{array}{|c|c|l|}\hline
\lambda&\mu& P_{[\lambda,\mu]}(a,b;q)  \\
\hline  
\emptyset&\emptyset &1 \\
\hline
(1)&\emptyset& a+b+a^2b^2
\\
\hline
(2)&\emptyset& 
a^2+b^2+a^4b^4
+(1+q)a b(1+a^2 b+a b^2)
\\
\hline
(1,1)&\emptyset&
q(a^2+b^2+a^4 b^4)
+(1+q)a b(1+a^2 b+a b^2)
\\
\hline
\emptyset&(1)&
1+a^2 b+a b^2
\\
\hline
\emptyset& (2)&
q(1+a^4 b^2+a^2 b^4)
+(1+q)a b(a+b+a^2 b^2)
\\
\hline
(1)&(1)&
(1+q+q^2)a^2b^2+q a b(a^2+b^2) +q(a+b)(1+a^3b^3)
\\
\hline
(1)&(2)& 
(1 + q + 2q^2 + q^3)a^2b^2(1 + a^2b + a b^2) 
+ q(1 + q)a b(a^2 + b^2 + a^4b^4) 
\\
&&
+ q^2(a + b +a^2 b^2(a^3 + b^3) + a^4 b^4(a^2 + b^2))
\\
\hline
\end{array}
\]
This polynomial is thought of an analogue of the Umemura polynomials
which arise from 
algebraic solutions of the Painlev\'e differential equations; 
{\it cf.} \cite{noou}.
\end{example}

\section{Verification of Proposition~\ref{prop:uc} }
\label{sect:ver}

Take an $(l+l'+2) \times (l+l'+2)$ matrix of the form:
\begin{align}
X &=(X_{a,b})_{1\leq   a,b \leq l+l'+2}  
\nonumber
\\
&=
\left(
\begin{array}{l | l | l }
-{t_i}^{-1}T_j(H_{\mu_{l'-a+1}  +a - 1 }) &
-{t_j}^{-1}T_i(H_{\mu_{l'-a+1}  +a - 1 }) &
T_{ij}(H_{\mu_{l'-a+1}  +a - b+2 })
\\
\hline
T_j(h_{\lambda_{a-l'-2}-a+2}) &
T_i(h_{\lambda_{a-l'-2}-a+2}) &
T_{ij}(h_{\lambda_{a-l'-2}-a+b})
\end{array}
\right)
\begin{array}{l}
\bigr\} \ {l'+1}\\
\bigr\}  \ l+1
\end{array} .
\\
& \nonumber
 \quad  \,   \,
\underbrace{     \qquad  \qquad  \qquad \qquad   \   \    }_{1} 
\underbrace{     \qquad  \qquad  \qquad \qquad    \  \    }_{1} 
\underbrace{     \qquad  \qquad  \qquad \quad     \          }_{l+l'}
\end{align}
Let $D=\det X$ and denote by 
$D[i_1,i_2,\ldots;j_1,j_2,\ldots]$
its minor determinant removing rows $\{i_a\}$
and columns $\{j_a\}$.
We put  $\lambda_0=k$ and $\mu_0=k'$.
\begin{lemma}
\label{lemma:D}
It holds that
\begin{subequations}
\begin{align}
(t_i-t_j)s_{[(k,\lambda),(k',\mu) ]}({\boldsymbol t})
&=
(t_it_j)^{l'+1} D,    \label{eq:claim1}
\\
T_{ij}(s_{[\lambda,\mu ]}({\boldsymbol t}) )
&=D[l'+1,l'+2;1,2],  \label{eq:claim2}
\\
T_i(s_{[(k,\lambda),\mu ]}({\boldsymbol t}) )
&=(-t_j)^{l'}D[l'+1;1],
\\
T_j(s_{[\lambda,(k',\mu) ]}({\boldsymbol t}) )
&=(-t_i)^{l'+1}D[l'+2;2],
\\
T_j(s_{[(k,\lambda),\mu ]}({\boldsymbol t}) )
&=(-t_i)^{l'}D[l'+1;2],
\\
T_i(s_{[\lambda,(k' ,\mu) ]}({\boldsymbol t}) )
&=(-t_j)^{l'+1}D[l'+2;1].    \label{eq:claim6}
\end{align}
\end{subequations}
\end{lemma}

\pf
Let us prove only (\ref{eq:claim1})
in the following;
the others (\ref{eq:claim2})--(\ref{eq:claim6}) can be verified in a similar manner.
It is easy to see that
\begin{subequations}  \label{subeq:dis}
\begin{eqnarray}
T_i(h_n) &=& h_n-t_i h_{n-1},  \label{eq:dis1}
\\
T_i(H_n) &=& H_n-{t_i}^{-1} H_{n-1}. \label{eq:dis2}
\end{eqnarray}
\end{subequations}

We shall apply 
elementary transformations 
successively
to
the row vector  
$\left(
h_n,h_{n+1}, \ldots,h_{n+r-1}
\right)$
of size $r=l+l'+2$ .
First we add the $b^{\rm th}$ 
column multiplied by $-t_i$
to the $(b+1)^{\rm th}$ column for $1\leq b \leq r-1$. 
We then obtain by  (\ref{eq:dis1}),
\[
\left(
h_n,T_{i}(h_{n+1}),T_{i}(h_{n+2}),  \ldots,T_{i}(h_{n+r-1})
\right).
\]
Secondly adding the $b^{\rm th}$ column multiplied by $-t_j$
to the $(b+1)^{\rm th}$ column for $2 \leq b \leq r-1$, 
we get
\[
\left(
h_{n},T_{i}(h_{n+1}),T_{ij}(h_{n+2}), \ldots,T_{ij}(h_{n+r-1})
\right).
\]
Adding the second column multiplied by $(t_i-t_j)^{-1}$
to the first column, 
we finally obtain the vector:
\[
\left(
(t_i-t_j)^{-1}T_{j}(h_{n+1}),T_{i}(h_{n+1}),T_{ij}(h_{n+2}), \ldots,T_{ij}(h_{n+r-1})
\right).
\]

By the same procedure as above,
the low vector
$\left(
H_n,H_{n-1}, \ldots,H_{n-r+1}
\right)$  is also 
converted to
\[
\left(
-(t_i-t_j)^{-1} t_j T_{j}(H_n),-t_iT_{i}(H_{n}),t_it_jT_{ij}(H_{n}), \ldots,t_it_jT_{ij}(H_{n-r+3})
\right),
\]
via (\ref{eq:dis2}).

Therefore,
remembering  (\ref{eq:def-of-uc-2}), 
we arrive at the expression (\ref{eq:claim1}).
\qed
\\

\noindent
{\it Proof of Proposition~\ref{prop:uc}.}  \
By the use of Jacobi's identity:
\[
D  D[l'+1,l'+2;1,2]=
D[l'+1;1]  D[l'+2;2]
- D[l'+1;2]D[l'+2;1],
\]
we see that (\ref{eq:bil-of-uc}) follows immediately from
Lemma~\ref{lemma:D}.
\qed

\newpage

 \appendix
\section{Reductions to $q$-Painlev\'e equations of types  $A_{2g+1}^{(1)}$ and $D_5^{(1)}$}

Recall that
the 
$q$-Painlev\'e equations of types  $A_{2g+1}^{(1)}$ and $D_5^{(1)}$ 
can be derived as 
reductions 
from the 
$q$-UC hierarchy;
see  \cite{t3} 
and \cite{tm}.
Accordingly, 
they can be derived
also from 
the lattice $q$-UC hierarchy,
as the latter hierarchy  
includes the former one;
see Remark~\ref{remark:q-uc}.
We verify 
that 
the equations of types  $A_{2g+1}^{(1)}$  and $D_5^{(1)}$
are
in fact 
similarity reductions of the lattice $q$-UC hierarchy together with 
periodic conditions of 
order $(g+1,g+1)$ and $(2,2)$,
respectively.
In this appendix,
we demonstrate how to obtain the
$q$-Painlev\'e equation  
only for the case of type $D_5^{(1)}$;
the other case 
is simpler, 
so it may be left to the reader;
{\it cf.}  \cite{t3}.

Let 
$I=\{1,2\}$ and $J=\{-1,-2\}$.
Suppose that 
$\sigma_{m,n}=\sigma_{m,n}({\boldsymbol t })$
is a solution 
of the lattice $q$-UC hierarchy (\ref{eq:l-q-uc}),
satisfying 
the periodic condition
$\sigma_{m,n}=\sigma_{m+2,n}= \sigma_{m,n+2}$
and the similarity condition
$\sigma_{m,n}(c {\boldsymbol t})=
c^{d_{m,n}} \sigma_{m,n}( {\boldsymbol t})$.
Here $d_{m,n}$ are constants balanced as  
$d_{m,n}+d_{m+1,n+1}=d_{m+1,n}+d_{m,n+1}$. 
Now let us introduce  the
function  $\rho_{m,n}(\alpha,\beta;x)$ in $x$,
equipped with
constant parameters $\alpha$ and $\beta$,  
defined by
$\rho_{m,n}(\alpha,\beta;x)=\sigma_{m,n}({\boldsymbol t})$
under the substitution 
${\boldsymbol t}=(t_1,t_2,t_{-1},t_{-2}) =(\alpha,\alpha^{-1},-q^{-1}\beta x, -q^{-1}\beta^{-1}x)$.
Let 
\begin{equation}
\begin{array}{ll}
\Phi_i^{(-)}(x)= \rho_{i,i}(\alpha,\beta;x ),
\quad
&\Phi_i^{(+)}(x)= \rho_{i,i}(q^{1/2}\alpha,q^{1/2}\beta;x ),
\\
\Psi_i^{(-)}(x)= \rho_{i,i+1}(\alpha,q^{1/2}\beta;q^{1/2}x ),
\quad
&\Psi_i^{(+)}(x)= \rho_{i,i+1}(q^{1/2}\alpha,\beta;q^{1/2}x ),
\end{array}
\end{equation}
for $i \in {\mathbb Z}/2{\mathbb Z}$.
As similar to the case of $E_6^{(1)}$
(see Section \ref{sect:sim}),
we therefore obtain,
from (\ref{eq:l-q-uc}) 
with the above constraints,
the following system of  
bilinear equations: 
\begin{subequations}    
                     \label{subeq:bil-qp6}
\begin{align}
&  
\alpha^{\pm 1}  q^{(d_{i,i}-d_{i,i+1})/2}
 \Phi_i^{(\pm)}(x)\Phi_{i+1}^{(\mp)}(x)
+
\beta^{\pm 1} x 
q^{(d_{i+1,i}-d_{i,i})/2} 
\Phi_i^{(\mp)}(x)\Phi_{i+1}^{(\pm)}(x)
\nonumber  \\
&
=
\left(\alpha^{\pm 1}+\beta^{\pm 1} x\right)
\Psi_i^{(\pm)}(q^{-1}x)\Psi_{i+1}^{(\mp)}(x),
\\  
&
\alpha^{\pm 1}  q^{(d_{i,i+1}-d_{i,i})/2}
\Psi_i^{(\pm)}(x)\Psi_{i+1}^{(\mp)}(x)
+
\left(q^{1/2}\beta \right)^{\mp 1} 
\left(q^{1/2}x \right) 
q^{(d_{i+1,i}-d_{i,i})/2}
\Psi_i^{(\mp)}(x)\Psi_{i+1}^{(\pm)}(x)
\nonumber  \\
&=
\left(\alpha^{\pm 1} + (q^{1/2}\beta)^{\mp 1} q^{1/2}x \right)
\Phi_i^{(\pm)}(x) \Phi_{i+1}^{(\mp)}(qx),
\end{align}
\end{subequations}
where $i \in {\mathbb Z}/2 {\mathbb Z}$.  
We take the variables
\begin{equation}
f(x)
=\frac{ \Phi_1^{(+)}(x)\Phi_{2}^{(-)}(x)  }{ \Phi_1^{(-)}(x)\Phi_{2}^{(+)}(x)   }  ,
\quad
g(x)
=\frac{ \Psi_1^{(+)}(x)\Psi_{2}^{(-)}(x)  }{ \Psi_1^{(-)}(x)\Psi_{2}^{(+)}(x)   } ,
\end{equation}
and let $\gamma=q^{(d_{1,1}-d_{1,2})/2}$  and   $\delta=q^{(d_{2,1}- d_{1,1})/2}$.
Hence it follows from (\ref{subeq:bil-qp6}) that
\begin{subequations}  \label{subeq:qp6}
\begin{align}
\overline{f}f
&= 
\frac{(g + \alpha^{-1} \beta^{-1} \gamma \delta x)(g +  \alpha  \beta \gamma^{-1} \delta^{-1}  q x)}
{(x g + \alpha\beta\gamma\delta)( q x g + \alpha^{-1}\beta^{-1}\gamma^{-1}\delta^{-1} )},
\\
\underline{g}g
&=
\frac{(f+\alpha^{-1}\beta \gamma^{-1}\delta x )(f+\alpha\beta^{-1} \gamma\delta^{-1} x )}
{(x f +\alpha \beta^{-1} \gamma^{-1}\delta  )( x f +\alpha^{-1} \beta \gamma \delta^{-1} )},
\end{align}
\end{subequations}
where the symbols $\overline{f}$ and $\underline{g}$
stand for $f(qx)$ and $g(q^{-1}x)$,
respectively.
This system
is equivalent to
the $q$-Painlev\'e equation of type $D_5^{(1)}$,
known as the 
$q$-Painlev\'e VI equation;
see \cite{js}.
\\

\small
\noindent 
{\it Acknowledgements.} \
The author wishes to thank 
Tetsu Masuda,
Masatoshi Noumi, 
Yasuhiro Ohta,
Tomoyuki Takenawa,
and 
Yasuhiko Yamada
for valuable discussions.
This work is partially supported by 
a fellowship of the Japan Society for the Promotion of Science (JSPS).


\noindent
Teruhisa Tsuda  \quad
\verb|tudateru@math.kyushu-u.ac.jp|
\\
Faculty of Mathematics, Kyushu University,
Hakozaki, Fukuoka 812-8581, Japan

\end{document}